\documentclass[aip, 
apl, amsmath,amssymb,
reprint]{revtex4-1}

\usepackage{graphicx}
\usepackage{dcolumn}
\usepackage{bm}

\usepackage[utf8]{inputenc}
\usepackage[T1]{fontenc}
\usepackage{mathptmx}
\usepackage{etoolbox}
\usepackage{textcomp}
\usepackage{gensymb}
\usepackage{xcolor}

\usepackage{soul}

\def\2D{$\mathrm{2D}$}
\def\3D{$\mathrm{3D}$}

\def\N2{$\mathrm{N}_2$}

\def\h2o{$\mathrm{H}_2\mathrm{O}$}
\def\co2{$\mathrm{CO}_2$}
\def\o2{$\mathrm{O}_2$}

\def\lfm{V\textsubscript{LFM}}

\begin{document}


\title{
    {Micro}fabricated Au and Au/graphene bilayer platelets for levitation experiments
} 



\author{Sunghyun Kim}
    \email{shdak6990@gmail.com}
    \affiliation{University of Maryland, College Park, MD, 20742, USA}
    \affiliation{Laboratory for Physical Sciences, 8050 Greenmead Dr., College Park, MD, 20740, USA}
 \author{Joyce E. Coppock}
    \email{jec@terpmail.umd.edu}
    \affiliation{University of Maryland, College Park, MD, 20742, USA}
    \affiliation{Laboratory for Physical Sciences, 8050 Greenmead Dr., College Park, MD, 20740, USA}
 \author{B. E. Kane}
    \email{bekane@umd.edu}
    \affiliation{Laboratory for Physical Sciences, 8050 Greenmead Dr., College Park, MD, 20740, USA}
    \affiliation{Joint Quantum Institute, University of Maryland, College Park, MD, 20742, USA}


\date{29 June 2026} 

\begin{abstract}
We describe a fabrication process for preparing liquid suspensions of micron-scale Au and Au/graphene bilayer platelets using thin-film deposition, optical lithography, ion milling, hydrofluoric acid (HF) substrate etching, and release from the substrate into a liquid suspension. 
Residual HF is removed through repeated centrifugation, decanting, and dilution cycles. 
The resulting suspension is characterized by electrospray deposition onto a secondary substrate, followed by electron and atomic force microscopy. 
The deposited platelets exhibit minimal aggregation, and the overall platelet yield reaches up to 30\% of the platelets originally patterned on the wafer. 
Lateral force microscopy further confirms that the Au/graphene bilayer remains intact throughout fabrication, release, and electrospray deposition. 
This process provides a practical route for preparing high-quality platelet suspensions for levitated nanoparticle experiments and other applications requiring suspensions of two-dimensional nanostructures.
\end{abstract}

\pacs{}

\maketitle 



Lithographic techniques have myriad applications in modern technology, enabling the fabrication of micro and nanoscale structures. In most cases, the fabricated objects remain attached to a substrate. Patterned objects, however, can also be released from a substrate to form liquid colloidal suspensions,\cite{Hernandez2007} potentially containing billions of microfabricated particles. One recent application of this idea is the fabrication of suspensions of nanoscale metallic objects with geometries optimized for plasmonic optical absorption.~\cite{Wi2011,KolarHofer2025} These engineered particles have been developed primarily for biomedical applications, demonstrating that lithographic fabrication can provide a versatile route for producing well-defined particle suspensions.


Another application of lithographically fabricated particle suspensions is as a source of nanoscale objects for levitation experiments using optical and quadrupole ion traps.~\cite{Gonzalez2021,Millen2020} While nearly spherical nanoparticles are readily available, engineered asymmetric particles are particularly attractive for rotational studies.~\cite{Ju2023,Zielinska2023,Kuhn2015,Pi2025} Recent work has demonstrated lithographically fabricated Si nanorods released from silicon-on-insulator wafers for levitation experiments.\cite{Pi2025}


For studies of the rotational and magnetic properties of levitated objects, graphene and other two-dimensional materials are of particular interest. Both rotational and magnetic measurements on graphene have previously been demonstrated,~\cite{Kane2010, Nagornykh2017} but {the source material relied on ultrasonic exfoliation of graphite}, a process that invariably produces irregularly shaped flakes with an uncertain number of layers. A fabrication method capable of producing single-layer graphene objects with well-defined geometries would therefore provide significant advantages. The commercial availability of wafer-scale single-layer graphene on Si substrates now makes lithographic fabrication of such structures feasible.

However, releasing and transferring micron-scale graphene structures from their supporting substrates while maintaining their integrity remains challenging because of the fragility of unsupported single-layer graphene. 
In this work, we describe a process for fabricating Au and Au/single-layer graphene platelets and producing suspensions of these particles for levitation experiments ultimately aimed at studying detached single-layer graphene. Au is selected as the carrier material because (1) it adheres well to graphene,\cite{Song2009,Torres2017,Zaborski2025} (2) it is resistant to the HF etches required to remove the underlying Si\o2 sacrificial layer, and (3) it can be readily removed from a levitated particle by heating and evaporation.~\cite{Coppock2021,Coppock2022}
To characterize the resulting suspensions, the particles are electrosprayed onto a secondary substrate and examined using electron microscopy (SEM) and atomic force microscopy (AFM), demonstrating successful release and preservation of the graphene layer.



\begin{figure}
    \centering
    \includegraphics[scale=0.5]{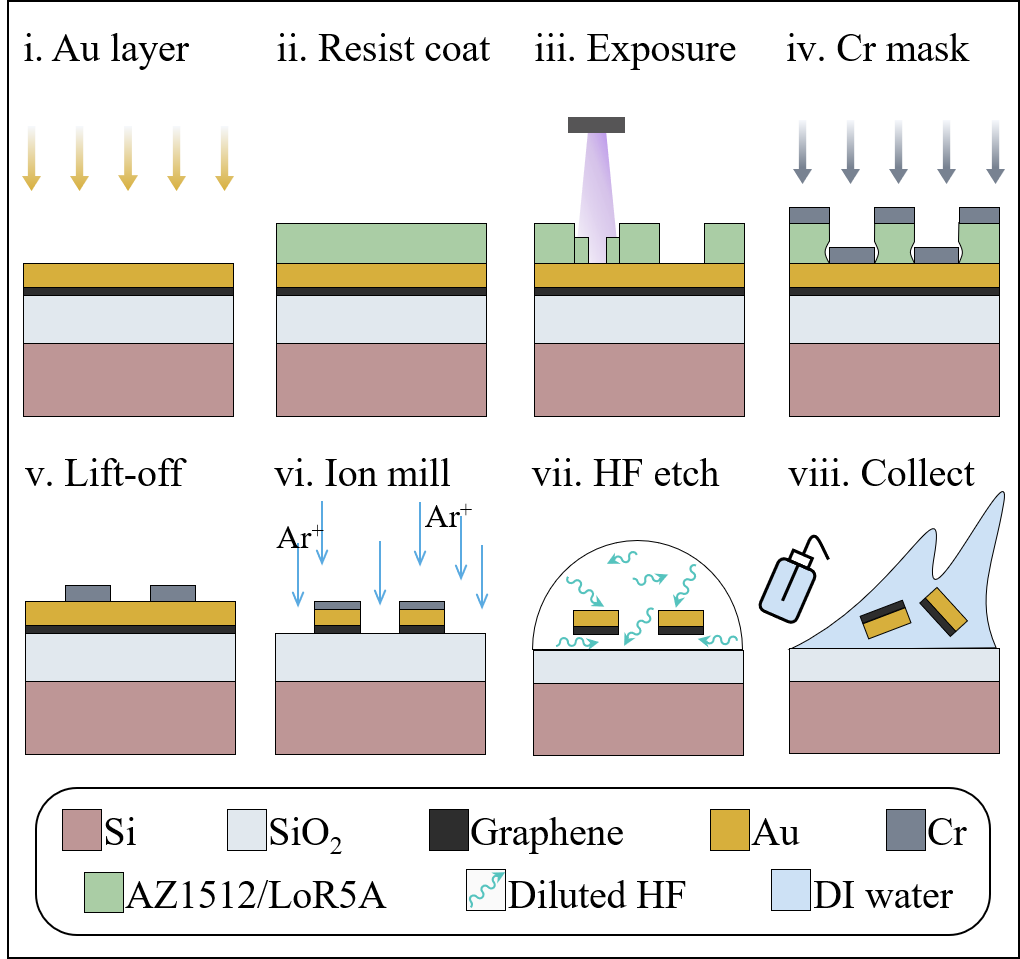}
    \caption{
    Fabrication workflow for Au/graphene platelets. The process consists of Au deposition on a graphene/Si\o2/Si wafer, photolithographic patterning, Cr hard-mask formation, ion milling, sacrificial Si\o2 etching, and platelet collection. The workflow for Au platelets is identical except for the absence of a graphene layer.}
\end{figure}
 
The platelet fabrication process is considered to produce stable platelet suspensions suitable for a broad range of laboratory experiments, including optical and quadrupole ion trapping systems used for levitation experiments.{~\cite{Wi2011,KolarHofer2025, Coppock2017,Nagornykh2015,Nagornykh2017,Coppock2021,Coppock2022,Coppock2024, Coppock2026, Kane2010, Kane2026}} First, the final product must be obtained as a stable suspension containing a sufficiently large number of platelets within a sub-milliliter volume. 
{Second, the suspension medium must be compatible with cleanroom processing and subsequent laboratory experiments, with any residual sacrificial-layer etchant (e.g., HF) sufficiently diluted to prevent corrosion of the electrospray plumbing and associated instrumentation.}
Third, the platelets should remain individually dispersed with minimal aggregation and surface contamination. Finally, for graphene-integrated samples, the fabrication process must preserve the integrity of the graphene {underlayer}.

Figure~1 summarizes the fabrication workflow developed for producing suspended Au/graphene and Au platelets. The two platelet types are fabricated using an identical process, differing only in the choice of starting substrate. Therefore, the Au/graphene platelet is used as the representative example throughout this paper unless otherwise specified.

A single-layer graphene on Si/SiO\textsubscript{2} wafer with a 285 nm SiO\textsubscript{2} layer is used as the starting substrate for the fabrication of Au/graphene platelets. \footnote{The graphene wafer is purchased from Grolltex (San Diego, CA, USA).} 
A 50 nm thick Au film is deposited onto the graphene surface by electron-beam evaporation using a CHA e-beam evaporation system at a deposition rate of $\sim$1 \AA/s ({Fig.~1}(i)). Platelet geometries are defined by optical lithography ({Fig.~1}(ii)--(iii)) with an exposure dose of 285 mJ/cm$^2$.\footnote{Optical lithography is performed by using a Heidelberg Instruments (Heidelberg, Germany) MLA150 direct-write system} A bilayer AZ1512/LoR5A photoresist is used for lift-off. The exposed photoresist is sequentially developed using AZ 1:1 and AZ 726 MIF, followed by rinsing with deionized (DI) water.
Next, a 12 nm thick Cr layer is deposited at a deposition rate of $\sim$1 \AA/s and patterned by lift-off to form Cr hard masks ({Fig.~1}(iv)--(v)). Lift-off is performed by overnight immersion in Remover PG at room temperature, followed by acetone spraying and rinsing with IPA and DI water.

The platelet patterns are then transferred into the Au/graphene stack by ion milling ({Fig.~1}(vi)). 
\footnote{Ion milling is performed at the NIST Center for Nanoscale Science and Technology (Gaithersburg, MD, USA).} 
During milling, the sample stage is rotated at 100 rpm to ensure uniform etching, while a built-in secondary ion monitor provides {\it in situ}, real-time monitoring of the milled elements for endpoint detection. Under the milling conditions used in this work, Au and Cr are etched at rates of about 2.8 and 0.6 \AA/s, respectively. To ensure complete removal of the Cr hard mask, the milling process is intentionally continued beyond the endpoint, thereby sacrificing $\sim$10 nm of the underlying Au layer.

\begin{figure}
    \centering
    \includegraphics[scale=0.45]{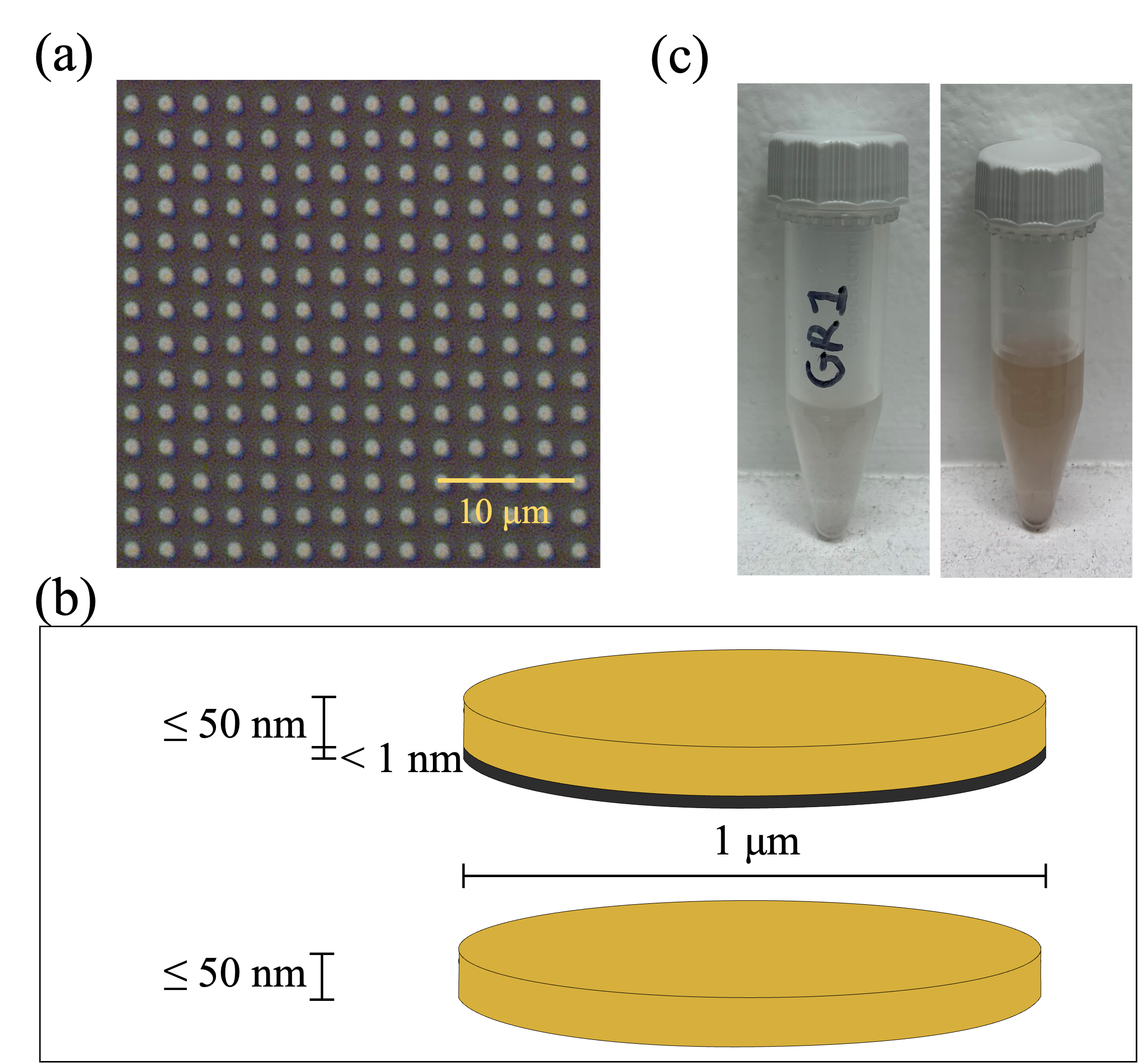}
    \caption{
    (a) Optical microscopy image of the patterned platelet array following ion milling.
    (b) Schematic illustrations of Au/graphene (top) and Au (bottom) platelets.
    (c) Photographs of the resulting Au/graphene (left) and Au (right) platelet suspensions after release, collection, centrifugation, and washing.}
    \label{fig:1}
\end{figure}

The processed wafer is inspected by optical microscopy after ion milling (Fig.~2(a)). Light-colored circular features on the darker SiO\textsubscript{2} background form a regular array across the wafer. The lithographic pattern is designed to yield about {$1.3\times 10^{9}$} Au/graphene platelets over the 100 mm wafer, each with a diameter of 1~$\mu$m and a thickness of $\leq$50 nm, as illustrated in Fig.~2(b). Optical microscopy of the as-received graphene layer indicates a grain size of $\geq$20~$\mu$m. Since the platelet diameter is 1~$\mu$m, each platelet is expected to contain a single graphene grain. 

Unlike conventional lift-off approaches, the present process employs the underlying SiO\textsubscript{2} layer as a sacrificial layer, enabling the release of free-standing Au/graphene platelets through selective HF etching. To minimize mechanical disturbance during release, a 400:1 dilution of HF in DI water is used with an etching time of 12 h (Fig.~1(vii)). {Under these gentle etching conditions, a substantial fraction of the platelets remains loosely adhered to the wafer after the etching process, rather than becoming suspended in the HF solution. In the present experiment, although the 30 ml of diluted HF solution completely covers the wafer, its nonuniform distribution during etching further results in incomplete underetching across the wafer.}

Since most of the platelets remain visibly on the wafer surface after HF etching, the diluted HF solution is first discarded before platelet collection. The platelets are then removed from the substrate by gentle DI water jetting using a squeeze bottle (Fig.~1(viii)), in which the lateral flow of DI water carries the released platelets away from the wafer. Approximately 20 ml of DI water is used for jetting, and the resulting platelet suspension is collected in a 50 ml PTFE centrifuge tube. 
Visual inspection of the wafer after the jetting process indicates that about half of the fabricated Au/graphene platelets have been removed from the substrate.

Then, 20 ml of a 2 mM aqueous (NH\textsubscript{4})\textsubscript{2}CO\textsubscript{3} solution prepared in IPA/DI water (1:3, v/v) is added, yielding a total suspension volume to 40 ml.~\cite{Coppock2021} The suspension is centrifuged at an RCF of $\sim$2600 g for 5 min and decanted to reduce the volume from 40 to 4 ml. Fresh (NH\textsubscript{4})\textsubscript{2}CO\textsubscript{3} solution is then added to restore the suspension volume back to 40 ml, followed by vortex mixing to wash the platelets. The centrifugation, decanting, and washing steps are repeated five times. 
The minute amount of residual HF carried into the suspension is neutralized through the repeated washing cycles with the (NH\textsubscript{4})\textsubscript{2}CO\textsubscript{3} solution.
After the final cycle, the suspension pH is confirmed to be $\sim$7 using pH test strips. Finally, the suspension is concentrated by one additional centrifugation and decanting step to a final volume of 2 ml.

The resulting platelet suspensions are transferred to 5 ml vials, as shown in the photographs in Fig.~2(c). Because only about half of the intended Au/graphene platelets are released and suspended in the present process, whereas the Au platelets are recovered much more efficiently, the Au/graphene suspension has a less intense orange color than the Au suspension.


To examine platelet distribution, surface morphology, aggregation, fabricated features, and other structural properties, 105~$\mu$l of the Au/graphene suspension is electrosprayed. 
The concentrated suspension is introduced into the ion trap system following an established loading protocol.~\cite{Coppock2017} 
As illustrated in Fig.~3(a), the suspension is continuously delivered through a Pt-coated syringe needle with an inner diameter of 100~$\mu$m at a flow rate of 1~$\mu$l/min using a syringe pump.~\cite{Kane2026} 
{The} (NH\textsubscript{4})\textsubscript{2}CO\textsubscript{3} {solution provides sufficient ionic conductivity for efficient charge transport during electrospray.}~\cite{Coppock2021}
A $+2500$ V DC bias is applied to the needle, causing electric charge to accumulate at the liquid meniscus and deform it into a stable Taylor cone. 
The suspension is subsequently emitted as a fine electrospray plume of charged droplets. A fraction of the charged droplets passes through the diaphragm aperture and enters the ion trapping chamber, 
{while another fraction is deposited on an aperture plate with a 75~$\mu$m-diameter central hole beneath the retainer (Fig.~3(a)).}~\cite{Kane2026} 
The aperture plate is subsequently collected and characterized by SEM and AFM.

\begin{figure}
    \centering
        \includegraphics[scale=0.44]{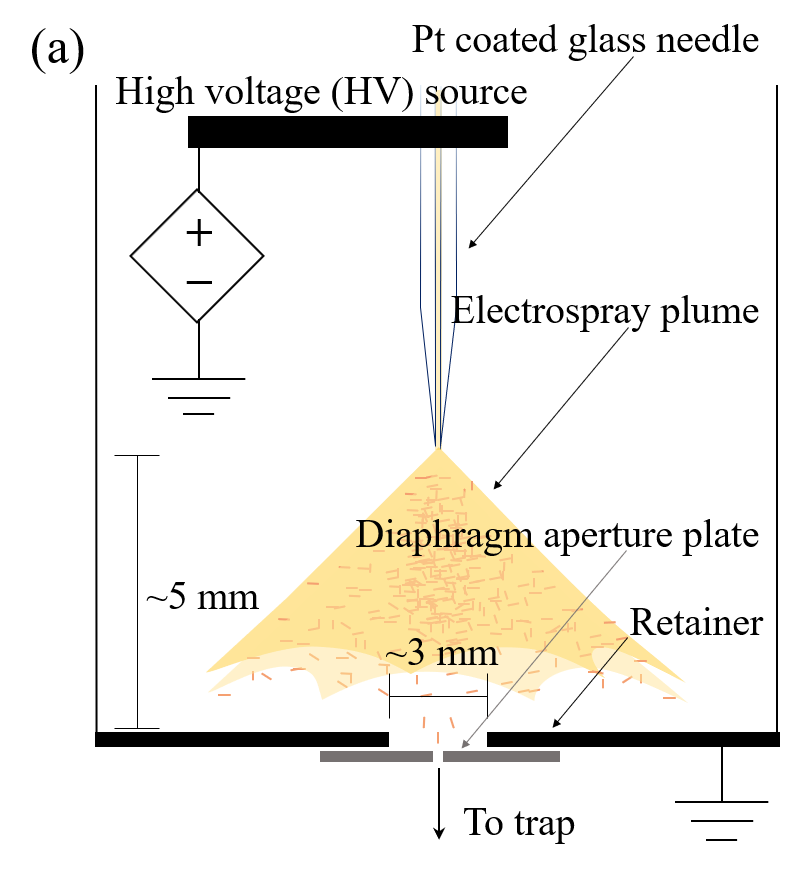}
        \includegraphics[scale=0.435]{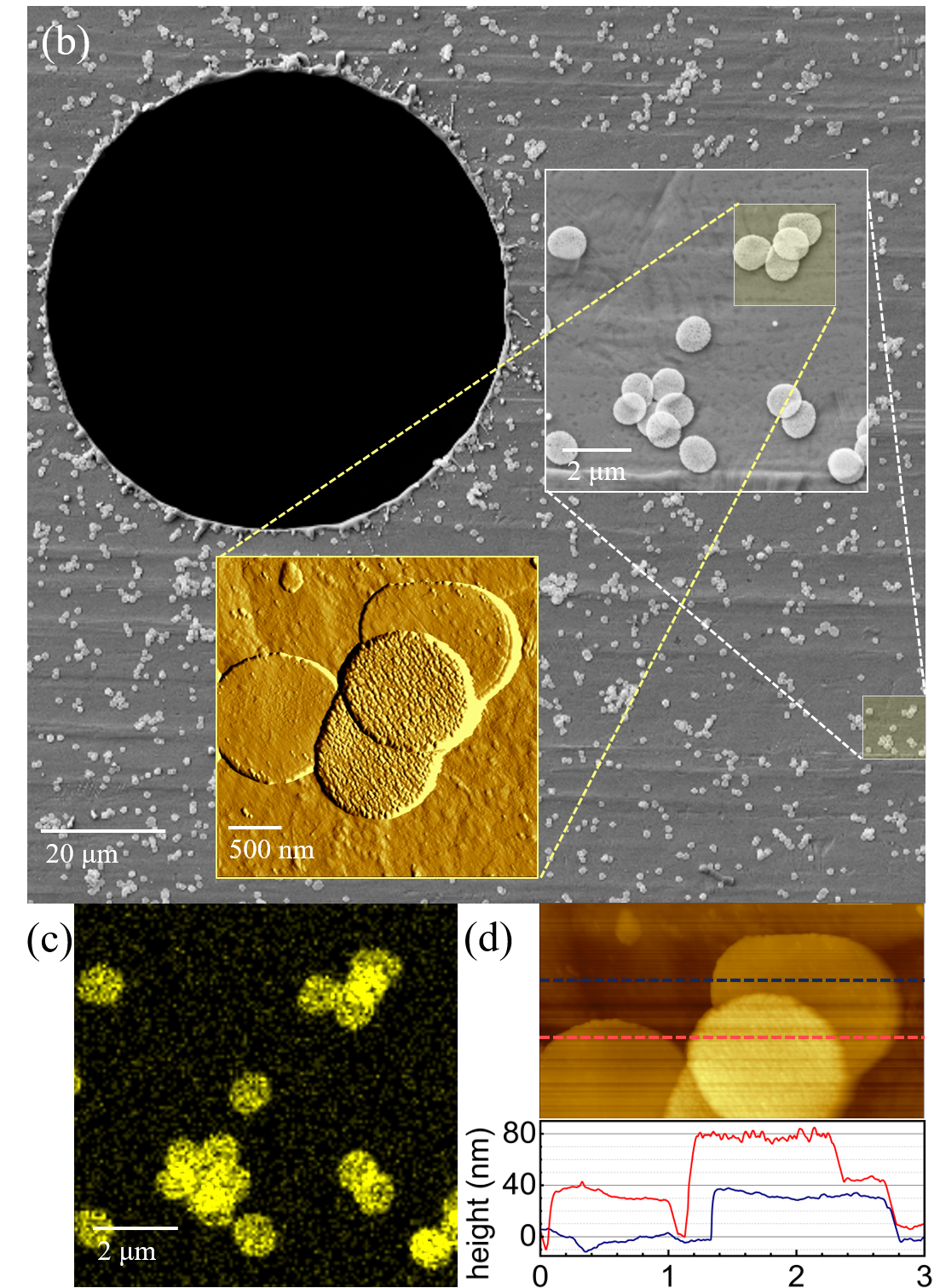}
    \caption{
    SEM and AFM characterization of electrosprayed platelets deposited on an aperture plate. 
    (a) Schematic illustration of the electrospray deposition. 
    (b) SEM image of electrosprayed Au/graphene platelets deposited near the aperture. The 75~$\mu$m aperture opening is visible in the upper left corner of the image. Insets show a magnified SEM image and a tapping-mode AFM deflection-error image acquired from the indicated subregions. 
    (c) EDS map of Au acquired from the SEM subregion in (b). 
    (d) AFM height image and corresponding line profiles extracted along the dashed lines. 
    }
\end{figure}

Prior to measuring the actual platelet density deposited near the aperture, the expected density is estimated for an ideal case. The total number of platelets patterned on the wafer is approximately $1.3\times10^{9}$, and the final suspension volume is 2 ml, yielding an ideal concentration of $\sim6.5\times10^{8}$ cts/ml. Since 105~$\mu$l of the suspension is electrosprayed in the present experiment, $\sim6.8\times10^{7}$ platelets are emitted through the Taylor cone.
Assuming an ideal Taylor-cone geometry with a half-angle of $45^\circ$ and a uniform angular distribution of emitted platelets, the electrospray plume intercepts a spherical surface area of approximately $2\pi(1-\cos45^\circ)(5~\mathrm{mm})^2 = 46~\mathrm{mm}^2$ at the aperture plate position (Fig.~3(a)), yielding an expected platelet density of approximately $1.5\times10^{6}$ cts/mm$^{2}$.

Figure~3(b) shows an SEM image acquired across the deposition region. The Au/graphene platelet density measured near the aperture is approximately $1.3\times10^{5}$ cts/mm$^{2}$, corresponding to $8.7\%$ of the ideal density. This reduction is attributed to losses during both the platelet fabrication process and the electrospray loading process. Because only approximately half of the intended Au/graphene platelets are released during the present fabrication process, the overall survival yield of the Au/graphene platelets cannot be determined reliably. In contrast, Au platelets fabricated using an otherwise identical process exhibit an overall survival yield of approximately $30\%$ from lithographic patterning to electrospray, suggesting that incomplete release of the Au/graphene platelets is presently the dominant source of particle loss.

Higher-resolution SEM and AFM are performed over $10\times10~\mu$m$^2$ and $3\times3~\mu$m$^2$ subregions, respectively, at the locations indicated in Fig.~3(b). The SEM images reveal disk-shaped platelets exhibiting two distinct surface morphologies, hereafter referred to as rough and smooth. Energy-dispersive x-ray spectroscopy (EDS) mapping (Fig.~3(c)) confirms that these structures are composed of Au. No apparent photoresist residues or other organic contaminants are observed on the platelet surfaces. Although platelet aggregates are occasionally present, the constituent platelets retain either the rough or smooth surface morphology.

The two surface morphologies are attributed to the opposite sides of the platelets. The rougher surface corresponds to the side exposed to Au deposition and ion milling, whereas the smoother surface corresponds to the side that remains in contact with the graphene/Si\o2 substrate during fabrication. Similar surface contrast is observed for both unreleased platelets on the wafer and released Au platelets, supporting this interpretation. Although EDS confirms the presence of Au, detection of graphene is limited by the weak signal from the atomically thin carbon layer and background carbon contamination.

To determine the platelet thickness, tapping-mode AFM is performed,    \footnote{Tapping-mode AFM employs a stiff-$k$ tip (BudgetSensors, Tap300AI-G, $k=40$ N/m).} and the thickness is obtained from the corresponding line profile (Fig.~3(d)). After background leveling, the measured thickness is $\sim$40 nm. This value is smaller than the nominal design thickness, 50 nm, which is attributed to intentional overmilling during removal of the Cr hard mask.

\begin{figure}
    \centering
    \includegraphics[scale=0.45]{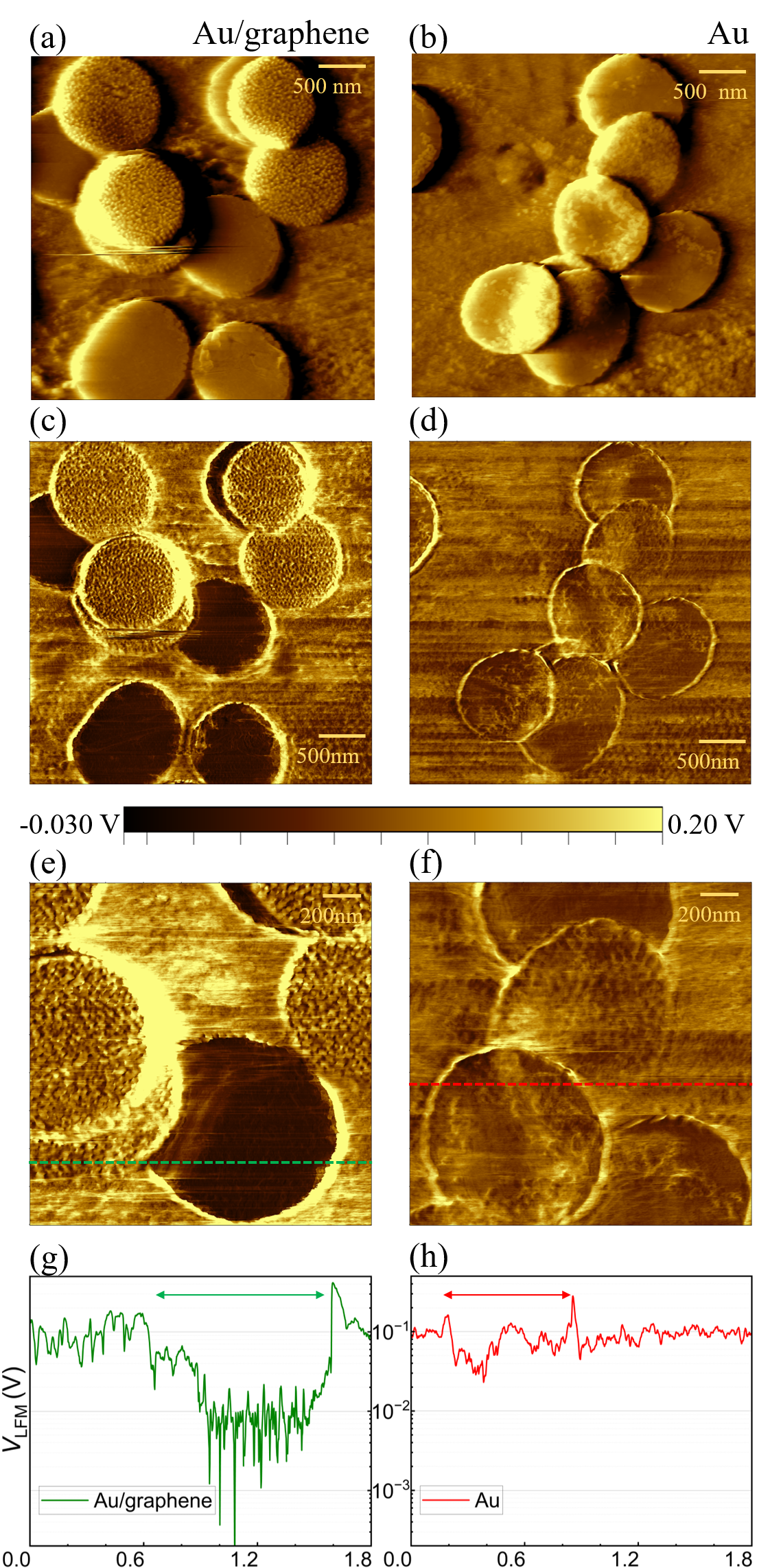}
   \caption{
    Comparison of Au/graphene (left column) and Au (right column) platelets. (a,b) Contact-mode AFM deflection-error images. 
    (c,d) LFM difference images calculated as $(\mathrm{Trace}-\mathrm{Retrace})/2$. 
    (e,f) Magnified LFM images acquired from  $1.8\times1.8~\mu\mathrm{m}^2$ subregions of (c) and (d). 
    (g,h) Cross-sectional profiles extracted along the dashed lines in (e) and (f), with arrows indicating the smooth-side regions. 
}
\end{figure}

The remaining question is whether the graphene layer survives the fabrication and electrospray without delamination from the Au film. Figure~4 compares Au/graphene and Au platelets after electrospray deposition. Contact mode AFM images (Fig.~4(a)--4(f)) reveal the surface morphologies of both the rough and smooth sides of the platelets. Because the graphene layer is expected to reside on the smooth side, the following discussion focuses on this surface.

As shown in Fig.~4(a), the smooth side of the Au/graphene platelet appears atomically flat and featureless over most of the scanned area. Only small chips and folded regions are observed near the platelet edges, which are attributed to repeated contact-mode AFM scanning under a positive normal load.~\cite{Chang2018} In contrast, overlapping Au platelets exhibit pronounced topographic variations on the smooth-side surface (Fig.~4(d)), producing a noticeably rougher morphology than that observed for Au/graphene platelets. Isolated Au platelets, however, remain relatively smooth. The absence of such topographic features on Au/graphene platelets suggests that the graphene {underlayer} helps preserve the surface morphology throughout fabrication and electrospray.

Because of its ultralow friction, graphene can be distinguished from Au using lateral force microscopy (LFM).~\cite{Kawai2016,Moon2025} To quantify the friction signal (\lfm), the half-width of the LFM loop, $(\mathrm{Trace}-\mathrm{Retrace})/2$, is extracted from trace and retrace LFM images.~\cite{Ogletree1996} The outer margins of the scanned images are cropped prior to analysis to improve contrast and eliminate edge artifacts unrelated to friction measurements. LFM measurements are performed in contact mode at a scan speed of approximately 10~$\mu$m/s.\footnote{Contact-mode AFM employs a soft-$k$ tip (Bruker, SNL-D, $k=0.06$ N/m)  under a positive normal load.} These scan conditions are chosen to provide reliable LFM signals while minimizing feedback-induced artifacts.~\cite{Riedo2003,kim2018} Rather than determine absolute friction forces, to compare friction contrast between Au and graphene, the \lfm\ values are reported in the native photodiode voltage units without lateral-force calibration.

The presence of graphene on the platelet surface is confirmed by the ultralow friction signal observed in the LFM measurements (Fig.~4(c) and~4(e)). Representative \lfm\ cross sections are shown in Fig.~4(g) and~4(h). Gaussian fits to friction-signal distributions obtained from three independent LFM scans yield \lfm\ values of $9.8\pm7.2$ mV for graphene (smooth side) and $85\pm38$ mV for Au (rough side). In contrast, Au platelets exhibit a similar \lfm\ value of $63\pm19$ mV on both the rough and smooth sides despite the differences in surface roughness. The persistence of the low-friction graphene signature after fabrication, release, collection, and electrospray demonstrates that the graphene layer remains firmly attached to the Au platelet throughout the entire process.

The present process is demonstrated using direct-write optical lithography to fabricate 1~$\mu$m-diameter platelets, but it is readily adaptable to alternative lithographic techniques, such as nanoimprint lithography,~\cite{Wi2011,KolarHofer2025} for producing substantially smaller particles. The present fabrication approach may also be combined with alternative particle-loading methods used in levitation experiments, including laser-induced acoustic desorption techniques,~\cite{Bykov2019,Gonzalez2021} thereby extending its applicability beyond electrospray injection. A long-term objective is the realization of levitated single-layer graphene by removing the Au support layer after trapping. Although this remains technically challenging, several possible approaches may enable its implementation. Even without Au removal, the fabricated Au/graphene bilayers provide a platform for studying graphene growth on nanoscale Au surfaces.~\cite{Coppock2026}

More broadly, the present fabrication and suspension techniques here can be extended to a wide range of material systems. Suitable lithographic and release processes may be applied to magnetic materials, eutectic alloys, refractory metals, semiconductor heterostructures, and multilayer structures composed of dissimilar materials.~\cite{Zhang2012,Rahman2026}  Such approaches enable the fabrication and suspension of nanoscale structures with well-defined dimensions, geometries, and material compositions. Furthermore, lithographic techniques allow the realization of anisotropic particle geometries that are difficult to achieve using conventional particle synthesis methods.~\cite{Kuhn2015,Pi2025} These capabilities may prove useful for future studies of rotational dynamics, controlled spinning motion, torque generation, and magnetic levitation experiments requiring orientation control.~\cite{Stickler2021}


\begin{acknowledgments}
This work is supported by the Laboratory for Physical Sciences, Contract \#H9823023C0086. {This work was performed in part at the NIST Center for Nanoscale Science and Technology.}
\end{acknowledgments}

\section*{Author Contributions}
\textbf{S.K.:} 
Investigation, Methodology, Validation, Formal analysis, Project administration, Visualization, Writing -- original draft, Writing -- review \& editing.
\textbf{J.E.C.:} Investigation, Methodology, Project administration, Writing -- review \& editing. 
\textbf{B.E.K.:} Conceptualization, Supervision, Writing -- original draft; Writing -- review \& editing.

\section*{Notes}
The authors have no conflicts to disclose. The authors declare no competing financial interest.

\bibliography{Lithography}

\end{document}